\documentclass[11pt]{article}

\usepackage[a4paper,margin=1in]{geometry}
\usepackage{amsfonts,amssymb,amsmath,amsthm}
\usepackage{graphicx}
\usepackage{psfrag}
\usepackage{calc}
\usepackage{epic}
\usepackage[english]{babel}
\usepackage{hyperref}
\usepackage{float}
\usepackage{latexsym}
\usepackage{color}
\usepackage{subfigure,dsfont}

\usepackage{hyperref}
\newcommand{\footremember}[2]{%
    \footnote{#2}
    \newcounter{#1}
    \setcounter{#1}{\value{footnote}}%
}
 
\title{Stability conditions for a discrete-time decentralised medium access algorithm
}
\author{%
  Seva Shneer\footremember{HW}{Heriot-Watt University, V.Shneer@hw.ac.uk}%
  \and Alexander Stolyar\footremember{UIUC}{University of Illinois at Urbana-Champaign, stolyar@illinois.edu}%
  }

\newtheorem{theorem}{Theorem}
\newtheorem{lemma}[theorem]{Lemma}

\newtheorem{definition}[theorem]{Definition}

\theoremstyle{remark}
\newtheorem{remark}[theorem]{Remark}

\newcommand{\be}{ \begin{equation}}
\newcommand{\ee}{\end{equation}}
\newcommand{\ben}{ \begin{equation*}}
\newcommand{\een}{\end{equation*}}

\newcommand{\beql}[1]{\begin{equation}\label{#1}}
\newcommand{\eeql}{\end{equation}}
\newcommand{\eqn}[1]{(\ref{#1})}

\def\E{{\mathbb E}}

\def\I{{\mathbb I}}

\begin{document}
\maketitle

\begin{abstract}
We consider a stochastic queueing system modelling the behaviour of a wireless network with nodes employing a discrete-time version of the standard decentralised medium access algorithm. The system is {\em unsaturated} -- each node receives an exogenous flow of packets at the rate $\lambda$ packets per time slot. Each packet takes one slot to transmit, but neighbouring nodes cannot transmit simultaneously.
The algorithm we study is {\em standard} in that: a node with empty queue does {\em not} compete for medium access; the access procedure by a node does {\em not} depend on its queue length, as long as it is non-zero. 
Two system topologies are considered, with nodes arranged in a circle and in a line.
We prove that, for either topology, the system is stochastically stable under condition $\lambda < 2/5$. This result is intuitive for the circle topology as the throughput each node receives in a saturated system (with infinite queues) is equal to the so-called {\em parking constant}, which is larger than $2/5$. (This fact, however, does not help to prove our result.)
The result is not intuitive at all for the line topology as in a saturated system some nodes receive a throughput lower than $2/5$.
\end{abstract}

\section{Introduction}

In this paper we consider a stochastic queueing model, primarily motivated by the MAC (Medium Access Control) algorithms in wireless networks.

Wireless devices are an indispensable part of everyday life for most of us. Wireless networks are used to connect devices at home and at businesses, to connect cell phones, sensors and even satellites. The sizes of wireless networks are growing at astonishing speeds, and with the development of Internet of Things will only continue to grow.

Wireless transmissions inevitably create noise and if many devices transmit simultaneously, the packets may be lost. There is therefore a need for a MAC  algorithm that will control which devices are going to transmit at a given time. In wireless networks there is a natural notion of neighbouring devices meaning that the transmissions generated by these devices interfere with each other. The model in this paper corresponds to those MAC protocols that prevent neighbours from transmitting simultaneously, thus making any collision and loss of packets impossible.

In this paper we focus on the {\em single-hop} wireless networks, where each generated packet only needs to be transmitted once, by its source node (where the packet was generated).  
We are interested in stability of a network, i.e. the ability of the nodes in the network to transmit all generated packets, without the packet queues building up to infinity. 
If an access algorithm guarantees stability for all packet arrival processes for which stability is in principle possible, such an algorithm is called maximally stable (or throughput-optimal).

The celebrated MaxWeight (or, BackPressure) algorithms have been introduced in \cite{Tassiulas1992}  and have been shown to be maximally stable. 
They have later been extended and generalised in a number of directions. Maximum stability is also known to be achieved in the single-hop case by a class of $\alpha$-fair algorithms (see \cite{Kelly1998, Mo2000, Roberts2000} for introduction of the fair-allocation concepts and \cite{Bonald2001, DeVeciana2001} for stability proofs).
All these algorithms are centralised, in that they need a central controller to know the states of the queues of all nodes and then to solve an optimisation problem to make every access control decision. Having such a central entity (and its ability to solve an optimisation problem for a vast number of devices) is not feasible in large networks. There is therefore a need for the decentralised algorithms where each node regulates its own access to the medium.

A well-known and widely used (most notably in IEEE 802.11) example of a decentralised MAC algorithm preventing collisions is the CSMA (Carrier Sense Multiple Access) protocol, where each node has a random back-off time before attempting to access the channel and if, on expiration of the back-off time, it does not sense any other node transmitting, the node will start its own transmission. Let us call this type of multiple access a {\em standard CSMA}, where `standard' refers to the following properties: (a) each node does not know (and does not try to explicitly learn) its neighbours, (b) the access procedure is same regardless of the node queue length, as long as it is non-zero, and (c) the node does not access the channel when its queue length is zero (i.e., no packets to transmit).

Most results on networks governed by the standard CSMA protocol assume that the system is saturated, i.e. each node is assumed to always have a packet to transmit (or, has infinite queue), and the question of interest is the throughput of the system, or an individual node. The typical assumption is that the service times, as well as back-off times, are Exponentially distributed, and the state of the system is fully described by the activity process tracking which nodes are currently transmitting. This Markov process turns out to be reversible and it thus has a product-form stationary distribution, providing values of throughputs of individual nodes (see \cite{Boorstyn1987, Durvy2009, Wang2005}).

In practice nodes generate packets at random time instances and thus do not always compete for medium access. This calls for the analysis of unsaturated systems, modelled by each node having its own exogenous arrivals of generated packets. Such unsaturated systems may {\em not} be approximated by saturated ones, or even bounded by them. For example,
 if one exchanges an unsaturated node for a saturated one, this will be detrimental to the performance of the node's neighbours but beneficial for the neighbours' neighbours (with whom the original node does not interfere). In other words, the random process describing such a system does {\em not} have the {\em monotonicity} property. Monotonicity, informally speaking, means that two versions of the process, with initial state of one dominating that of the other, can be coupled so that this dominance relation persists at all times. The absence of monotonicity substantially complicates the process analysis, including establishing stability conditions. 
It is also impractical in applications to keep a device accessing the network if it does not have packets to transmit. (Hence property (c) of a standard CSMA is important.) The analysis of general unsaturated networks is extremely difficult, because, primarily due to lack of monotonicity, the queue dynamics and transmission schedule of any node depends on the states of all other nodes, in a very complicated manner. 

Very few results are known for unsaturated systems (\cite{Cecchi2014, Ven2010}). In \cite{Ven2010} the authors consider a continuous-time model and study the question of stability. 
They demonstrate with an example that the condition that the arrival intensity for each node is smaller than the throughput of the node in the saturated system is not sufficient for stability. This means, in particular, that standard CSMA does {\em not} achieve maximum stability. 

The absence of maximal stability of the standard CSMA protocol led to the development, starting with \cite{JW2010}, of queue-based algorithms where the distribution of the back-off duration of a node depends on the state of its queue. In particular, an algorithm of this type was proposed in \cite{Shah2012} for a continuous-time model, which guarantees maximal stability for single-hop systems on any graph. However, despite providing maximal stability, the queue-based algorithms are difficult to implement and are known to lead to high delays. (Hence, property (b) of a standard CSMA is important).


In this paper, our goal is to characterise the performance of a standard CSMA algorithm. 
We would like to stress again that standard CSMA algorithms are important, because they are decentralised (and therefore easy to implement), because implementing a queue-based scheme leads to long delays, and because it is 
impractical in the unsaturated situation to keep all nodes active at all times.
We will focus on two simple single-hop systems, consisting of $N$ nodes arranged in a circle or in a line. The systems operate in discrete time.  To model standard CSMA in discrete, slotted time, we will assume that at the beginning of each time slot the nodes are given access priorities, forming a permutation of numbers $1,\ldots,N$,
picked independently (across time slots), uniformly at random from all possible permutations. The node with the highest priority transmits in this slot if its queue is non-empty. The node with the second-highest priority transmits in the slot if its queue is non-empty and if none of its neighbour nodes is transmitting. And so on until all nodes are checked in their priority order. All transmission times are equal to $1$, so at the beginning of the next time slot no transmission is ongoing, and the medium access process is repeated independently. Note that this algorithm is easy to implement in a decentralised fashion, with an arbitrarily small loss in throughput (see below for a precise algorithm).

Our main result concerns the homogeneous system, where the arrival rates into each node are equal. Non-homogeneous systems are notoriously difficult and even the simpler random access algorithm ALOHA does not allow for the exact description of the stability region (see, e.g., \cite{Bonald2004, Bordenave2008, Borst2008, Szpanowski1994}). 

The maximum stability for a standard CSMA system on a circle or on a line is given by $\lambda < \lfloor N/2\rfloor /N$, where $\lambda$ is 
the average number of new generated packets per unit of time. (If $N$ is even, the condition is simply $\lambda < 1/2$.) This is not achievable by the CSMA protocol on a circle. Indeed, if all nodes have non-empty queues, then the process of transmission scheduling by the nodes in a time slot is equivalent to the discrete {\em parking problem} (see, e.g., \cite{Flory1939, Page1959}), and there are known expressions for the expected total number of transmissions per time slot (see \cite{Shneer2017} for expressions adopted to CSMA setting). On a circle, if all nodes have non-empty queues, then by symmetry the expected number of transmissions per slot is the same for all nodes and is equal to the so-called parking constant (sometimes referred to as jamming density), which we denote by $c_p=c_p(N)$. Therefore, if $\lambda$ exceeds $c_p$, the system is unstable. The parking constant $c_p$ for a circle of $N$ nodes is equal to $1/2$ if $N=4$, $2/5$ if $N=5$ and it decreases over even values of $N$ and increases over odd values of $N$ to the same limit $1/2(1-e^{-2}) \approx 0.4323$.

On the other hand, each non-empty node always transmits if its priority is higher than the priorities of both its neighbours, which happens with probability $1/3$. Therefore $\lambda < 1/3$ trivially leads to stability. One of the goals of our work is to study how close the stability region of the standard CSMA algorithm on a circle is to $\lambda < c_p$ (the best achievable for this algorithm).

The {\bf main result} of the paper for the {\bf circle topology} is that, under the standard CSMA, $\lambda < 2/5$ leads to stability for a system with $N\ge 4$. We conjecture that $\lambda < c_p$ is sufficient for stability, but this is not proved.

It appears to be a lot more difficult to conjecture the stability condition for a system on a line. If all nodes have non-empty queues, then the process of transmission scheduling in a slot is again equivalent to the discrete parking problem, but in this situation the expected number of transmissions differs from node to node. For instance, the nodes on the edges will have transmission probabilities larger than $1/2$ and tending to $1-e^{-1} \approx 0.6321$ as $N\to\infty$, but the nodes right next to them will have much lower transmission probabilities tending to $e^{-1} \approx 0.3679$ (see \cite{Shneer2017}).

Our {\bf main result} for the {\bf line topology} is that, just like for the circle system, $\lambda < 2/5$ leads to stability for a system with $N \ge 4$. This result is not intuitive at all, given that in the saturated system the transmission probability of the second node from the edge is strictly less than $2/5$ (see above). The result also stresses once again that the saturated system does not provide a bound for the unsaturated one, that it is unreasonable to make all nodes (including empty) to compete for the transmission at all times, and that
it is important to study unsaturated systems.

Our main results are proved using the fluid limit technique. The main difficulty in the proofs consists in dealing with situations when a number of queues in the fluid limit are at zero. One has to study rather complicated structure of the occupancy and activation processes of the neighbourhood of non-zero queues. We believe that the technique we developed may be used to study the behaviour of CSMA and similar decentralised algorithms on more complicated topologies.

We also provide a relatively short direct proof of stability for both topologies under condition $\lambda < 3/8$. This is weaker than our main results but we find that the proof provides a good insight into the behaviour of the model. The lack of efficiency of the random-access algorithm compared to the centralised one on a circle or a line comes from the fact that at each time slot a number of pairs of neighbouring nodes will not transmit. Therefore we consider the Lyapunov function $\sum_i (Q_i+Q_{i+1})^2$, where $Q_i$ is the queue length at node $i$. 
Another advantage of this approach is that it provides some sufficient conditions for stability of inhomogeneous systems (with different arrival rates at the nodes). 

The paper is structured as follows. We introduce the model and notation in Section \ref{sec:model}. In Section \ref{sec:results} we formulate and prove auxiliary results and formulate our main results. Section \ref{sec-proof-main} is devoted to the proof of the main results. We provide some conjectures and discuss our further research plans in Section \ref{sec:conclusion}.

\section{Model and notation} \label{sec:model}

Assume that we have $N$ nodes (transmitters) on a circle or on a line, and for any node we are going to say that the nodes to its immediate left and right are its neighbours. More formally, if we consider the circle topology, we will denote the neighbourhood of node $i$ as
\begin{eqnarray*}
\mathcal{N}_c(i) = 
\begin{cases}
\{i-1,i+1\} \quad \text{for} \quad i=2,\ldots, N-1, \\
\{N, 2\} \quad \text{for} \quad i=1, \\
\{N-1, 1\} \quad \text{for} \quad i=N,
\end{cases}
\end{eqnarray*}
and if we consider the line topology, we will denote the neighbourhood of node $i$ as
\begin{eqnarray*}
\mathcal{N}_l(i) = 
\begin{cases}
\{i-1,i+1\} \quad \text{for} \quad i=2,\ldots, N-1, \\
\{2\} \quad \text{for} \quad i=1, \\
\{N-1\} \quad \text{for} \quad i=N. 
\end{cases}
\end{eqnarray*}

Assume that each node has unlimited buffer space to queue packets waiting for transmission. Time is slotted.
At the beginning of each time slot, first
new transmissions are initiated, and then the arrivals of new packets (messages) occur. Every message transmission time is equal to $1$, so any transmission is completed before the beginning of the next time slot.

Denote by $Q_i(n)$ the queue length at node $i$ at time $n$. Denote by $\xi_i(n)$ the number of new arrivals into node $i$ at time $n$. We assume that $\xi_i(n)$ are all i.i.d. with $\E \xi_i(n) = \lambda$.

We will assume that neighbours cannot transmit in the same time slot. The competition for medium access is as follows. At the beginning of each time slot $n$, an allocation of priorities $\{U_1(n),\ldots, U_N(n)\}$ for nodes is chosen uniformly at random from all possible permutations of the set $\{1,\ldots,N\}$. The node with the highest priority $1$ will transmit if its queue is not empty. The further procedure is defined by induction: the node with the next-highest priority will transmit if its queue is not empty and if no node in its neighbourhood is already transmitting. At each time slot this procedure is applied until the set of all nodes is exhausted, and this procedure is repeated independently at each time slot. Upon completion of its transmission, the packet leaves the system. 

Denote by $D_i(n)$ the number
of packets transmitted by node $i$ in time slot $n$. Note that $D_i(n)$ can only take values $0$ and $1$.
Then the evolution of queue $i$ length is as follows:
$$
Q_i(n+1) = Q_i(n) - D_i(n) + \xi_i(n) .
$$

\begin{remark}
Note that the discrete time medium access procedure described above may be implemented in a decentralised way with a negligible loss in efficiency. Indeed, assume that at the beginning of each time slot $n$, every node with a non-empty queue (every node $i$ such that $Q_i(n) > 0$) generates a random variable $W_i(n)$ uniformly in $(0,\varepsilon)$. We assume that the random variables $\{W_i(n)\}$ are independent in $i$ and $n$. These random variables serve as back-off times for transmitters: node $i$ will wait exactly $W_i(n)$ and at that time, if no node in its neighbourhood is transmitting, will start its transmission. If at least one of the nodes in the neighbourhood of node $i$ is transmitting at time $n+W_i(n)$, then node $i$ stays silent for the duration of the time slot. As we can see, only an $\varepsilon$ proportion of each time slot needs to be devoted to this competition, and $\varepsilon$ may be made arbitrarily small. 
\end{remark}

Some basic notation we use throughout the paper: the norm of a vector $q=(q_1, \ldots, q_N)$ is $\| q \| \doteq \sum_i |q_i|$; $\lfloor x \rfloor$ is the largest integer not greater than $x$; $\mathbb{I}(A)$ denotes the indicator of an event $A$.

\section{Results} \label{sec:results}

In this section we present the results of our paper. We start (in Section \ref{subsec:parking}) by formulating and proving some properties of expected numbers of transmissions on intervals which are used in our proofs and which are also interesting in their own right. We then proceed (in Section \ref{subsec:reshuffling}) to prove maximum stability achievable by our algorithm (standard CSMA), for a system where nodes change their locations at random at the beginning of every time slot. In Section \ref{subsec:38} we turn our attention back to the original system and prove stability for both circle and line topologies in the case $\lambda <3/8$ using a one-step Lyapunov function. Finally, in Section \ref{subsec:main} we present our main results stating that the original system for both circle and line topologies is stable provided $\lambda < 2/5$. The proof of the main results is based on fluid limits and is given in Section \ref{sec-proof-main}.

\subsection{Some properties of parking constants for a segment} \label{subsec:parking}

Denote by $L_k$ the expected number of departures from a network consisting of $k$ nodes on a line which are assumed to be non-empty. Note again that the transmission initiation process on a set of non-empty nodes is equivalent to the discrete-time parking problem, therefore $L_k$ may be thought of as the expected number of cars parked on the segment, and $L_k/k$ -- as the parking constant, or jamming density. We note here that the original papers (see, e.g. \cite{Flory1939,Page1959}) on the parking problem are devoted to the expected amount of sites occupied by the parked cars (and each car needs two sites), which is closely related to but slightly different from the number of expected transmissions in our case. Formulas adapted to the CSMA situation are provided in \cite{Shneer2017} and we are going to refer to this paper hereinafter.

\begin{lemma} \label{lemma:monotonicity_E}
$L_n/n$ is non-increasing for $n \ge 3$.
\end{lemma}

{\bf Proof.} We need to show that $\frac{L_n}{n} \le \frac{L_{n-1}}{n-1}$ for all $n\ge 3$, which is equivalent to $n(L_n-L_{n-1}) - L_n \le 0$ for $n\ge 4$. Note that, using the expression for $L_n$ from \cite[Corollary 2]{Shneer2017},
\begin{align*}
L_n-L_{n-1} & = & \sum_{k=1}^n (-1)^{k+1} \frac{2^{k-1}}{k!}(n-k+1) - \sum_{k=1}^{n-1} (-1)^{k+1} \frac{2^{k-1}}{k!}(n-k)
\\
& = & \sum_{k=1}^{n-1} (-1)^{k+1} \frac{2^{k-1}}{k!} + (-1)^{n+1} \frac{2^{n-1}}{n!} = \sum_{k=1}^{n} (-1)^{k+1} \frac{2^{k-1}}{k!},
\end{align*}
and hence
\begin{align*}
n(L_n-L_{n-1}) - L_n & = n \sum_{k=1}^{n} (-1)^{k+1} \frac{2^{k-1}}{k!} - \sum_{k=1}^n (-1)^{k+1} \frac{2^{k-1}}{k!}(n-k+1) \\ & = \sum_{k=1}^{n} (-1)^{k+1} \frac{2^{k-1}}{k!}(k-1) = \sum_{k=2}^n (-1)^{k+1} c_k,
\end{align*}
where we denoted $c_k = \frac{2^{k-1}}{k!}(k-1)$. It may be checked directly that the sum on the RHS of the last expression is negative for $n=4$ and $n=5$. Also, for any $n=2m$, the sum is smaller than the sum for $n=2m-1$, so it is sufficient to consider odd values of $n$. And the sums for odd values of $n\ge 5$ are negative as $c_{2m+1} \le c_{2m}$ for all $m \ge 3$.
\qed

\begin{lemma} \label{lemma:extra_competition}
Denote by $L_{k:m}$ the expected number of departures from $k$ first nodes in a network consisting of $m$ non-empty nodes. Then
\begin{equation} \label{eq:monotonicity1}
L_{k:m} \le L_k
\end{equation}
for any $k \le m$.
\end{lemma} 

The statement of the lemma may be described in words as "by adding interference, we reduce the expected number of departures".

{\bf Proof.}
We use induction. The statement is trivial for $m=1$. Assume that the statement is correct for all $m \le M-1$ and write
$$
L_{k:M} = \sum_{i=1}^k \frac{1}{M} (1+L_{i-2} + L_{k-i-1:M-i-1}) + \frac{1}{M} L_{k-1} + \sum_{i=k+2}^M \frac{1}{M} L_{k:i-2},
$$
with appropriate conventions.

We can also write
$$
L_k = \sum_{i=1}^k \frac{1}{M} (1+L_{i-2} + L_{k-i-1}) + \sum_{i=k+1}^M \frac{1}{M} L_{k},
$$
by adding $M-k$ dummy nodes to the right of node $k$. The statement is then easily seen by comparing the two formulas.
\qed

Formula \eqref{eq:monotonicity1} immediately implies

\begin{equation} \label{eq:monotonicity2}
L_{k+m} \le L_k + L_m
\end{equation}
for any $k$ and $m$. This can be described as "by stacking two non-empty pieces together, we make things worse in terms of average number of departures". Or, again, "added interference reduces expected number of departures".

\subsection{Stability of a system with reshuffling} \label{subsec:reshuffling}

We present here an artificial system which is not our primary interest and for which stability results are immediate. Unlike the system we investigate, this system features global, rather than local, interaction between nodes and its study is much simpler. 

Using Lemma \ref{lemma:monotonicity_E}, we easily obtain that condition $\lambda < \min\{L_N/N, 1/2\}$ implies $\lambda < L_k/k$ for all $k \le N$.

Denote by $C_k$ the expected number of departures from a network of $k$ non-empty nodes on a circle. 
For all $k\ge 1$, $C_k \le L_k$. Indeed, this is trivial for $k\le 2$. For $k\ge 3$, obviously, 
$C_k = L_{k-3} + 1$, and using  \eqn{eq:monotonicity2} it is easy to check that $L_{k-3}+1 \le L_k$.
Using $C_N \le L_N$, we can easily check that $\lambda < C_N/N$ implies $\lambda < \min\{L_N/N, 1/2\}$, and then implies 
$\lambda < L_k/k$ for all $k < N$.

This means that, {\em under condition $\lambda < \min\{L_N/N, 1/2\}$ for the line topology or $\lambda < C_N/N$ for the circle topology, on any non-empty segment of any length (including all nodes being non-empty), the expected number of arrivals in a time slot is smaller than the expected number of departures.}

For either topology, define the following system, which we will refer to as the {\em system with reshuffling}. In fact, we define three versions of it; in all of them arrivals happen according to the same rule as before. Version 1: at the beginning of each slot,
{\em all} system queues are reshuffled uniformly at random; i.e. queue locations are determined by a permutation of their indices, chosen uniformly at random. Version 2: at the beginning of each slot, all empty queues stay where they are, but {\em all non-empty} queues are reshuffled uniformly at random. Version 3: at the beginning of each slot, all empty queues stay where they are, but non-empty queues are reshuffled uniformly at random {\em within each non-empty segment.}

Given the property of non-empty segments derived above, we have the following result.

\begin{theorem}
For the circle topology, the system with reshuffling (any of the three versions) is stable if $\lambda < C_N/N$.
For the line topology, the system with reshuffling (any of the three versions) is stable if $\lambda < \min\{L_N/N, 1/2\}$.
\end{theorem}

\subsection{Stability in the case $\lambda <3/8$} \label{subsec:38}

In this section we prove that stability of the original system in either topology is guaranteed if $\lambda < 3/8$. This is weaker than our main results (also, the following theorem requires the second moment of the arrival process to be finite, which is not needed in our main result). However, we think that the proof sheds some light on the behaviour of the system as it is based on the Lyapunov function that penalises the cases when pairs of neighbouring nodes do not transmit - exactly what makes a random access inefficient on a circle or a line segment. The following theorem also has a simple generalisation leading to sufficient stability conditions for inhomogeneous arrival processes - something our main result does not allow.

\begin{theorem} \label{thm:38}
Let $\xi$ be a random variable equal in distribution to $\xi_i(n)$ for all $i$ and $n$. If $\lambda < 3/8$ and $\E(\xi^2) < \infty$, then the system in either topology is stable. 
\end{theorem}

{\bf Proof.} We use the standard Foster-Lyapunov technique. Let us start with the circle topology. Consider a function $L: \mathbb{R}_+^N \to \mathbb{R}_+$ given by
$$
L(x) = \sum_{i=1}^N (x_i + x_{i+1})^2,
$$
with the convention $x_{N+1} = x_1$.

It is sufficient to show that if $\lambda < 3/8$, then there exist $K < \infty$ and $\varepsilon > 0$ such that 
\begin{equation} \label{eq:Lyapunov_drift}
\E\left(L\left(Q(1)\right) - L\left(Q(0)\right)|Q(0) = Q\right) < -\varepsilon
\end{equation}
for all initial conditions $Q(0) = (Q_1(0),\ldots, Q_N(0)) = (Q_1,\ldots, Q_N) = Q$ with $\sum\limits_{i=1}^N Q_i \ge K$.

To show this, write
\begin{align} \label{eq:Lyapunov_estimate}
& \E\left(L\left(Q(1)\right) - L\left(Q(0)\right)|Q(0) = Q\right) \nonumber
\\ & = 
\sum_{i=1}^N \E\left((Q_i + Q_{i+1} + \xi_i + \xi_{i+1} - D_i - D_{i+1})^2 - (Q_i + Q_{i+1})^2\right) \nonumber \\
& \le 2 \sum_{i=1}^N (Q_i + Q_{i+1})(2\lambda - \E(D_i) - \E(D_{i+1}))
\\ & + \sum_{i=1}^N \E(\xi_i + \xi_{i+1})^2 + \sum_{i=1}^N \E(D_i + D_{i+1})^2, \nonumber
\end{align}
where for ease of notation we wrote $\xi_k$ for $\xi_k(0)$ and $D_k$ for $D_k(0)$. Note that in the last expression
$$
\sum_{i=1}^N \E(\xi_i + \xi_{i+1})^2 + \sum_{i=1}^N \E(D_i + D_{i+1})^2 \le 2\sum_{i=1}^N \E(\xi_i^2) + 2N\lambda^2 + 4N.
$$
Denote the constant on the RHS of the last expression by $C < \infty$. Let us now bound the first sum on the RHS of \eqref{eq:Lyapunov_estimate}. In order to do this, rewrite it as
$$
\sum_{i=1}^N (Q_i + Q_{i+1})(2\lambda - \E(D_i) - \E(D_{i+1})) = \sum_{i=1}^N Q_i (4\lambda - \E(D_{i-1}) - 2 \E(D_i) - \E(D_{i+1})).
$$
Now we consider all possible options for the states of the nearest neighbours of a non-empty node $i$. 

If $Q_{i-1}=Q_{i+1} = 0$, then
$$
\E(D_{i-1}) + 2 \E(D_i) + \E(D_{i+1}) = 2.
$$

If $Q_{i-1}=1$ and $Q_{i+1}=0$ (or the other way around), then
\begin{align*}
& \E(D_{i-1}) + 2 \E(D_i) + \E(D_{i+1}) = \E(D_{i-1}) + 2 \E(D_i) \\ 
& = (\E(D_{i-1}) + \E(D_i)) + \E(D_i) = 1 + E(D_i) \ge 3/2,
\end{align*}
as in this case exactly one of nodes $i-1$ and $i$ is transmitting, and with a probability at least $1/2$ it is node $i$.

Finally, if $Q_{i-1}=Q_{i+1} = 1$, then
$$
\E(D_{i-1}) + 2 \E(D_i) + \E(D_{i+1}) = (\E(D_{i-1}) + \E(D_i)) + (\E(D_i) + \E(D_{i+1})) \ge 3/2,
$$
where we use the fact that $\E(D_{i-1}) + \E(D_i) \ge 3/4$ for any two neighbouring non-empty nodes $i-1$ and $i$. To see this, note that the probability that neither node $i-1$ nor node $i$ is transmitting is bounded from above by the probability of the event that node $i-1$ has a priority lower than that of $i-2$ and node $i$ has a priority lower than that of $i+1$ (an easy observation shows that the complement of this intersection implies that either node $i-1$ or node $i$ will transmit), which is equal to $1/4$.

Combining these estimates, we can write
$$
\sum_{i=1}^N Q_i (4\lambda - \E(D_{i-1}) - 2 \E(D_i) - \E(D_{i+1})) \le (4\lambda - 3/2)\sum_{i=1}^N Q_i,
$$
and if $\lambda < 3/8$, then the RHS of \eqref{eq:Lyapunov_estimate} is does not exceed $-\varepsilon$, provided $\sum_{i=1}^N Q_i \ge K = (C+\varepsilon)/(2(3/2-4\lambda))$. This completes our proof for the circle topology.

The proof of Theorem \ref{thm:38} for the line topology follows the same lines and is based on the Lyapunov function
$$
L(\bar{x}) = \sum_{i=1}^{N-1} (x_i + x_{i+1})^2.
$$
Its expected drift in one step may be bounded from above by
\begin{align*}
& Q_1 (2\lambda - \E(D_{1}) - \E(D_2)) + \sum_{i=2}^{N-1} Q_i (4\lambda - \E(D_{i-1}) - 2 \E(D_i) - \E(D_{i+1})) 
\\ & +Q_N (2\lambda - \E(D_{N-1}) - \E(D_N))+ C.
\end{align*}
Now if $Q_1>0$, then $\E(D_{1}) + \E(D_2) = 1$ and the same holds for $Q_N$. The exact same arguments as used in the proof for the circle topology are applied to the rest of the drift, and we conclude that it is negative provided $\lambda < 3/8$. \qed

\begin{remark}
From the proof of Theorem \ref{thm:38} we can also see that for system with inhomogeneous arrivals, if we denote by $\lambda_i = \E(\xi_i(n))$, then
$$
\lambda_{i-1} + 2 \lambda_i + \lambda_{i+1} < 3/2
$$
is sufficient for stability. The above is guaranteed, for instance, by
$$
\lambda_i + \lambda_{i+1} < 3/4
$$
for every $i$.
\end{remark}

\subsection{Main results} \label{subsec:main}

The following lemma shows that for a circle of length $N \ge 4$, the parking constant is at least $2/5$.

\begin{lemma} \label{lemma:25}
For any $N \ge 4$, $C_N/N \ge 2/5$.
\end{lemma}

The proof of this lemma is presented in the Appendix.

Our main result for the system on a circle is the following Theorem~\ref{thm:main}. (Its proof is given in Section~\ref{sec-proof-main}.)
\begin{theorem} 
\label{thm:main}
The system with the circle topology and $N \ge 4$ is stable if $\lambda < 2/5$.
\end{theorem}

\begin{remark}
\label{rem-for-main}
Note that the system we consider is trivially stable under the condition $\lambda < C_N/N$, {\em if all nodes compete for transmission at all times, even when they are empty}. However, we want to emphasise that this fact does not imply Theorem~\ref{thm:main} and does not help to prove it. Recall that our main interest is the model of a standard CSMA, under which each node is ``completely unaware'' of its surrounding and competes for the channel when and only when it has packets to transmit. The difference between the models when all nodes are competing all the time and when empty nodes remain silent, is even more evident from our next result for the line topology.
\end{remark}

The main result for the line system has the same form as for the circle. (Its proof is also in Section~\ref{sec-proof-main}.)

\begin{theorem} \label{thm:main_line}
The system with the line topology and $N \ge 4$ is stable if $\lambda < 2/5$.
\end{theorem}

\begin{remark}
This result is more surprising than that for a circle as on a line, if all nodes have non-empty queues, their transmission probabilities are different. In fact, the transmission probability of the second node (or node $N-1$) is equal to $3/8$ if $N=4$, $11/30$ if $N=5$ and can be shown to decrease over even values of $N$ and increase over odd values of $N$ to the same limit $1-e^{-1} \approx 0.3679$ (see \cite{Shneer2017}), thus never exceeding $2/5$. This demonstrates that in order to obtain correct stability conditions, it is not sufficient to consider the throughputs of nodes in saturation.
\end{remark}

\section{Proof of main results}
\label{sec-proof-main}

\subsection{Fluid limits}

We will prove Theorem~\ref{thm:main} and Theorem~\ref{thm:main_line} using the fluid limit technique \cite{RybSt92, Dai95, St95}. 
(For the application of the fluid limit technique to {\em discrete time} processes, see e.g. \cite{CS2001}.)
This section gives definitions and preliminaries that will be needed in both proofs.

Denote by $\mathcal{N}= \{1,\ldots,N\}$ the set of all nodes, and by $Q(n) = (Q_i(n), ~i\in \mathcal{N})$ the queue length vector. The Markov process that we consider, $Q(n), ~n=0,1,\ldots$, is a discrete time countable irreducible Markov chain. To apply the fluid limit technique, consider a sequence of processes $Q^r(\cdot)$ with increasing initial state, $\|Q^r(0)\|=r \uparrow \infty$, is fixed. (Here and below, the upper index $r$, which is the norm of initial state, is used as a process label.)
To establish the process stability (positive recurrence) it suffices to prove that for some fixed integer $T>0$ any such sequence of processes is such that 
\beql{eq-fluid-key}
\E \frac{1}{r} \|Q^r(rT)\| \to 0, ~~r\to\infty.
\end{equation}
First, we will view process $Q^r(\cdot)$ (and all other system processes) in continuous time $t\ge 0$, by adopting the convention $Q^r(t) = Q^r(\lfloor t \rfloor)$, $t\ge 0$.
Define the associated processes as follows: $F_i^r(t)$ is the total number of exogenous arrivals in node $i$ by time $t$;
$H_i^r(t)$ is the total number of departures from (transmissions at) node $i$ by time $t$. We will also need an additional family of processes, $G_B^r(t)$, whose definition requires some details, given next.

Denote by $s$ a system {\em occupancy state} (at a given time), $s=(s_i, ~i\in \mathcal{N})$, where $s_i$ is 1 [resp. 0] if node $i$ is occupied [resp. empty]. There is, of course, only a finite number $2^N$ of possible $s$. At time $n$ the occupancy state is
given by $s_i = \I\{Q_i(n)>0\}$. Denote by $u=(u_i, ~i\in \mathcal{N})$ a {\em ranking realisation} (at a given time),
where $u_i \in \mathcal{N}$ is the ranking that node $i$ receives.
In other words, $u\in \mathcal{U}$, where $\mathcal{U}_N$ is the set of permutations of $(1, \ldots,N)$.
Denote by $d$ a {\em transmission realisation} (at a given time), $d=(d_i, ~i\in \mathcal{N})$,  $d_i$ is 1 [resp. 0] if node $i$ transmits [resp. does not transmit]. Clearly, at any time, $d=\phi(s,u)$ is a deterministic function of $s$ and $u$.
Denote by $\Psi$ the set of possible values of a triple $(s,u,d)$. Then, for $B \subseteq \Psi$, denote
by $G_B^r(t)$ the total number of time slots, up to and including time $t$, in which $(s,u,d) \in B$. 
(For example, $B$ can be a subset -- or, ``event'' -- like $\{Q_i>0 ~\mbox{and node $i+1$ transmits}\}$.) 
To simplify notation, we often write $G_{\sigma}^r(t)$ to mean $G_{s=\sigma}^r(t)$.

To summarize, the processes that we will consider are $Q_i^r(\cdot), F_i^r(\cdot), H_i^r(\cdot)$ for $i\in\mathcal{N}$,
and $G^r_B(\cdot)$ for $B\in \Psi$.

Recall that the processes with different index $r$ have different fixed initial states. 
We will construct all these processes (for all $r$) on a common probability space as follows.
The arrival process $F^r_i(\cdot)$ for a given $i$ (and for all $r$) is driven by an independent i.i.d. sequence of random variables $\xi_i(1), \xi_i(2), \ldots$, with finite mean $\E\xi_i(1)=\lambda$; namely, 
$$
F^r_i(t) = \sum_{1 \le n \le \lfloor t \rfloor} \xi_i(n).
$$
The processes that drive transmissions (for all $r$) are as follows. 
For each $s$ we define an independent i.i.d. driving sequence of random elements $u_s(\ell), ~\ell=1,2, \ldots$, 
distributed randomly uniformly in $\mathcal{U}_N$. When the system is in occupancy state $s$ for the $\ell$th time, the ranking realisation $u_s(\ell)$ is taken, which uniquely determines the transmission realisation $d=\phi(s,u_s(\ell))$.
Clearly, this collection of driving processes defines the entire process uniquely for each initial state.

Denote by $\Theta$ the set of all possible pairs $(u,d)$. For a given $s$, denote by
$\{P_s(u,d), ~(u,d)\in \Theta\},$ the probability distribution of a random outcome of $(u,d)$ when the system occupancy state is $s$.
In other words, $P_s(u,d)$ the probability distribution on $\Theta$, 
induced by $u$ distributed uniformly in $\mathcal{U}_N$, and (deterministic) mapping $d=\phi(s,u)$; clearly, 
$$
P_s(u,d) = \frac{1}{N!} \I\{d=\phi(s,u)\}.
$$
We use notation
$$
P_s(A) = \sum_{(u,d)\in A} P_s(u,d), ~~A \subseteq \Theta,
$$
for the probabilities of events, and $P_s(A|A')$ for conditional probabilities.

Note that the driving processes are such that the following FSLLN properties hold.
With probability 1, 
\beql{fslln1}
\frac{1}{r} F^r_i(rt) \to \lambda t,  ~\mbox{u.o.c.}, \forall i \in \mathcal{N},
\end{equation}
\beql{fslln2}
\frac{1}{r} \sum_{\ell \le r\tau} \I\{(u_s(\ell),\phi(s,u_s(\ell)) \in A \} \to P_s(A) \tau, ~\mbox{u.o.c.}, ~\forall A \subseteq \Theta, ~\forall s.
\end{equation}

For each $r$ we define fluid-scaled processes:
$$
q^r_i(t) = \frac{1}{r} Q^r_i(rt), ~~ f^r_i(t) = \frac{1}{r} F^r_i(rt), ~~ h^r_i(t) = \frac{1}{r} H^r_i(rt),
$$
$$
g^r_B(t) = \frac{1}{r} G^r_B(rt).
$$

\begin{definition}
\label{def-fsp}
A collection of deterministic functions 
$$\chi=[(q_i(\cdot), f_i(\cdot), h_i(\cdot), i\in \mathcal{N}), (g_B(\cdot), B \subseteq \Psi)]
$$ is called a {\em fluid limit} 
if there exists a subsequence of $r$ and the corresponding sequence of the (scaled) process realisations
$\chi^r=[(q^r_i(\cdot), f^r_i(\cdot), h^r_i(\cdot), i\in \mathcal{N}), (g^r_B(\cdot), B \subseteq \Psi)]$ such that,
as $r\to\infty$ along this subsequence,
\eqn{fslln1} and \eqn{fslln2} hold, and
$$
\chi^r \to \chi, ~~\mbox{u.o.c. for each component function}.
$$
\end{definition}

The following lemma, describing basic properties of fluid limits, is standard (see, e.g. \cite{RybSt92, Dai95, St95}), and so is its proof, which we will omit.

\begin{lemma}
\label{lem-basic-fsp-prop}
Any fluid limit $\chi$ has the following basic properties.\\
(i) All component functions are Lipschitz continuous. Therefore, a.e. ~ w.r.t. Lebesgue measure, all component functions have derivatives; time points, where this holds are caller {\em regular}.\\
(ii) $\|q(0)\| =1$.\\
(iii) $q_i(t) = q_i(0) + f_i(t) - h_i(t), ~~\forall i$.\\
(iv) $f_i(t) = \lambda t, ~t\ge 0, ~~\forall i.$\\
(v) $h_i(t) = g_{\{d_i=1\}}(t), ~t\ge 0, ~~\forall i.$\\
(vi) For any occupancy state $\sigma$ and any $A \subseteq \Theta$,
$$
g_{\{s=\sigma, (u,d) \in A\}}(t) = P_{\sigma}(A) g_{\sigma}(t).
$$
\end{lemma}

For a regular time point $t$, it will be convenient to use notation
$$
\pi_t(B) = (d/dt) g_B(t);
$$
and we sometimes will write $\pi_t(\sigma)$ to mean $\pi_t(s=\sigma)$.
Clearly, $\pi_t(\cdot)$ is a probability distribution on $\Psi$; $\pi_t(B|B')$ denotes conditional probabilities.

The following lemma, describing convergence to fluid limits, is also standard (see, again, \cite{RybSt92, Dai95, St95}) and we omit its proof as well.

\begin{lemma}
\label{lem-fluid-limit}
The sequence of fluid-scaled {\em processes} 
$$
\chi^r=[(q^r_i(\cdot), f^r_i(\cdot), h^r_i(\cdot), i\in \mathcal{N}), (g^r_B(\cdot), B \subseteq \Psi)
$$ 
is such that the following holds w.p.1. For any subsequence of $r$ there is a further subsequence, along which
$$
\chi^r \to \chi, ~~\mbox{u.o.c. for each component function},
$$
where $\chi$ is a fluid limit.
\end{lemma}

To prove Theorem~\ref{thm:main} and Theorem~\ref{thm:main_line}, it suffices to show that for some fixed $T>0$ and any fluid limit $\chi$,
\beql{eq-fluid-key2}
\|q(t)\| = 0, ~~\forall t\ge T.
\end{equation}
(The family of
random variables $\|q^r(T)\| \equiv (1/r) \|Q^r(rT)\|$ is easily seen to be uniformly integrable. This, 
along with \eqn{eq-fluid-key2} and Lemma~\ref{lem-fluid-limit}, implies \eqn{eq-fluid-key}.)

\subsection{Auxiliary definitions and facts}

\subsubsection{Ranking and conditional ranking realisation constructions}
\label{sec-ranking-alg}

Consider a set of $m$ nodes. They can be arranged in a circle or in a line. (For the definition that follows it is only required that a neighbour
relation between the nodes is defined.) For a given occupancy state $s$ the random process which determines the nodes' ranking realisation -- and the transmission realisation as well -- is as follows. The occupied nodes are initialised as {\em active} and idle nodes as {\em inactive}.
The ranking, from $1$ to $m$, is assigned to nodes in the order of being {\em chosen} (or, picked).
The first node is chosen uniformly at random among all $m$ nodes; if this node is active and occupied, it transmits and then deactivates itself and its neighbours; if this node is inactive, no further action is taken. The second node is chosen among the $m-1$ nodes not chosen before, and the action is taken according to the same rule as for the first node. And so on, until all nodes are chosen -- this produces the ranking and transmission realisations of all nodes.

Let two nodes $i$ and $j$ be fixed. Later on we will need to generate the ranking (and transmission) realisations, conditioned on the event $u_i < u_j$. It is easy to check that this can be done via the following construction.
The process, producing the ranking/transmission realisation, is the same as described above,
with the following modification. As long as both nodes $i$ and $j$ are not chosen yet, 
when the next node is being chosen,
nodes $i$ and $j$ are considered as a pair, which is picked with the probability twice greater than that for a single node.
If/when the pair $(i,j)$ is chosen, the node $i$ is picked. After that, the process runs as usual without modifications.

\subsubsection{Properties of occupied segments of nodes}

\begin{definition}
\label{def-segment}
For given occupancy state $s$,
a set of $k$ consecutive occupied nodes, bordering on idle nodes (or no nodes) on both sides, is called 
an {\em occupied segment} of length $k$. Without loss of generality, for each $k$, we can and will consider 
the occupied segment consisting of nodes $1,\ldots,k$, with nodes $N$ (in the case of a circle) and $k+1$ being idle,
and will denote it $[1,k]$.
\end{definition}

We are interested in the transmission probabilities of nodes within one occupied segment (and one time slot). Clearly, these probabilities
depend only on the random {\em mutual} ranking of the nodes within this segment. The mutual ranking $u'=(u'_1,\ldots,u'_k)\in \mathcal{U}_k$ 
is determined by total ranking $u$:  $u'_i < u'_j$ if and only if $u_i < u_j$.
The distribution of the mutual ranking $u'$ is uniform on the permutations set $\mathcal{U}_k$.
Formally speaking, we are interested in the probabilities $P_s(A)$ in the case when $s$ is such that
$[1,k]$ is an occupied segment, and $A$ depends only on the mutual ranking and transmissions within this segment. For this reason, the corresponding probabilities (with a slight abuse of notation) are denoted by $P_k(A)$.

\begin{definition}
\label{def-friend}
Nodes $i$ and $j$ within same occupied segment are called {\em friends} [resp. {\em foes}] if
$|j-i|$ is even [resp. odd].
\end{definition}

The following lemma means that if we add node $k+1$ or a pair of nodes $(k+1,k+2)$ to an occupied segment $[1,k]$,
this addition can only increase the transmission probabilities of the nodes in $[1,k]$ that are friends of node $k+1$,
and can only decrease the transmission probabilities of the nodes in $[1,k]$ that are foes of $k+1$.

\begin{lemma}
\label{lem-friend}
Consider an occupied segment $[1,k]$, and a subset $L$ of its nodes which are all mutual friends.
Consider also the larger occupied segments $[1,k+1]$ and $[1,k+2]$. Then, if node $k+1$ is a friend
[resp. foe] of the nodes in $L$, we have
\beql{eq-add1}
P_k(d_i=1, ~i\in L) ~\le ~\mbox{[resp. $\ge$]}~ P_{k+1}(d_i=1, ~i\in L),
\end{equation}
\beql{eq-add2}
P_k(d_i=1, ~i\in L) ~\le ~\mbox{[resp. $\ge$]}~ P_{k+2}(d_i=1, ~i\in L).
\end{equation}
\end{lemma}

{\bf Proof.} Let us prove \eqn{eq-add1}. (The proof of \eqn{eq-add2} is completely analogous.)
We will prove the following (stronger) statement. The ranking and transmission realisations of the segments
$[1,k]$ and $[1,k+1]$ can be coupled (constructed on a common probability space), so that for any $i=1,\ldots,k$:
($d_i =1$ in $[1,k]$-segment) implies ($d_i =1$ in $[1,k+1]$-segment) when $i$ is a friend of $k+1$;
and ($d_i =0$ in $[1,k]$-segment) implies ($d_i =0$ in $[1,k+1]$-segment) when $i$ is a foe of $k+1$.
This property is proved by induction on $k$. The statement is trivially true for $k=0$, if we adopt the convention that this case corresponds to the ``empty'' segment -- it has no nodes and, therefore, no friends or foes of any other node.
This is the base of induction. Suppose, the statement is true for $k=0,1,\ldots, m-1$. Let us prove it for $k=m$.
We are constructing the ranking and transmission realisations, jointly for the segments $[1,m]$ and $[1,m+1]$.
Segment $[1,m+1]$ initialised to have all nodes occupied and active; segment $[1,m]$ initialised to have all nodes occupied and active, but it is augmented by a ``dummy'' node $m+1$, which is initialised to be inactive.
The process for ranking/transmission realisation runs for $m+1$ nodes, jointly in both systems.
We stop this process when either (a) all nodes are inactive or (b) node $m+1$ is chosen. In the case (a), the transmission realisation for nodes $1,\ldots,m$ in both systems is same. In the case (b), there are two sub-cases: (b.1) node $m$ was already inactive and (b.2) node $m$ was active. In the case (b.1) the transmission of node $m+1$ in the $[1,m+1]$-system and its non-transmission in the $[1,m]$-system will not affect nodes $1,\ldots,m$ in either system, and therefore the rest of the ranking/transmission realisations in both systems can obviously be coupled to be the same.
The remaining case to consider is (b.2). In the $[1,m+1]$-system, the node $m$ -- a foe of $m+1$ -- is deactivated, and will not transmit; in the $[1,m]$-system the node $m$ remains active, and therefore may eventually transmit; so, the property we are proving holds for node $m$. The remaining active nodes in both systems are same, except the right-most occupied segment (if we consider only {\em active} occupied nodes) in the $[1,m]$-system has the additional node $m$ on the right. (As a degenerate special case, this segment may consist of the single node $m$ in the $[1,m]$-system
and be ``empty'' in the $[1,m+1]$-system.) This right-most segment has length at most $m$ in the $[1,m]$-system.
It remains to apply the induction hypothesis to complete the construction of coupling.
$\Box$

\begin{lemma}
\label{lem-edge-triv}
For any occupied segment of length $k\ge 1$
$$
P_k(d_1=1) \ge 1/2.
$$
\end{lemma}

{\bf Proof.} 
$
P_k(d_1=1) \ge P_k(u_1 < u_2) = 1/2
$.
$\Box$

\begin{lemma}
\label{lem-edge}
For any occupied segment of length $k\ge 2$
$$
P_k(d_1=1) \ge P_k(d_2=1).
$$
That is, an end node of a segment has the transmission probability not smaller than that of its neighbour.
\end{lemma}


{\bf Proof.} Clearly, $P_k(d_1=1~\mbox{or}~d_2=1) = 1$, and events $\{d_1=1\}$ and $\{d_2=1\}$ are mutually exclusive.
So, $P_k(d_1=1)+P_k(d_2=1) = 1$. And $P_k(d_1=1) \ge 1/2$ (Lemma~\ref{lem-edge-triv}).
$\Box$

\subsubsection{Additional properties of fluid limits}

First, note some standard properties that hold at a regular point of a fluid limit: if $q_i(t) = 0$, then necessarily $q'_i(t) = 0$, and then 
$h'_i(t) = g'_{\{d_i=1\}}(t) = \pi_t(d_i=1) = \lambda$; if $q_i(t) > 0$, then necessarily  
$\pi_t(s_i=1) = 1$.

\begin{lemma}
\label{lem-two-at-the-edge}
Consider a regular point $t$ such that $\pi_t(d_{i-1}=1)>0, q_{i}(t)>0, q_{i+1}(t)>0$.
Then,
$$
\pi_t(d_{i+1}=1~|~ d_{i-1}=1) \ge 1/2.
$$
\end{lemma}

{\bf Proof.} From $q_{i}(t)>0, q_{i+1}(t)>0$ we know that $\pi_t(s_i=1, ~s_{i+1}=1)=1$.
Observe that
$$
\pi_t(d_{i+1}=1, d_{i-1}=1) \ge \pi_t(d_{i+1}=1, d_{i-1}=1, u_{i+1} < u_{i+2}) = 
$$
$$
\pi_t(s_{i+1}=1, d_{i+1}=1, d_{i-1}=1, u_{i+1} < u_{i+2}) =
\pi_t(d_{i-1}=1, u_{i+1} < u_{i+2}),
$$
where we used $\pi_t(s_{i+1}=1)=1$ and the fact that $(s_{i+1}=1, d_{i-1}=1, u_{i+1} < u_{i+2})$ implies $d_{i+1}=1$.
Therefore, to prove the lemma it suffices to prove 
\beql{eq-111}
\pi_t(d_{i-1}=1, u_{i+1} < u_{i+2}) \ge (1/2) \pi_t(d_{i-1}=1),
\end{equation}
which is what we proceed to do.

We have
\begin{multline*}
\pi_t(d_{i-1}=1, u_{i+1} < u_{i+2})  = \sum_{\sigma} \pi_t(d_{i-1}=1, u_{i+1} < u_{i+2}, s=\sigma) 
\\ = \sum_{\sigma} \pi_t(\sigma) P_{\sigma}(d_{i-1}=1, u_{i+1} < u_{i+2}) 
\\ = \sum_{\sigma} \pi_t(\sigma) P_{\sigma}(u_{i+1} < u_{i+2}) P_{\sigma}(d_{i-1}=1~|~ u_{i+1} < u_{i+2})
\\ = (1/2) \sum_{\sigma} \pi_t(\sigma) P_{\sigma}(d_{i-1}=1~|~ u_{i+1} < u_{i+2})
\\ = (1/2) \sum_{\sigma: \sigma_{i-1}=\sigma_{i}=\sigma_{i+1}=1} \pi_t(\sigma) P_{\sigma}(d_{i-1}=1~|~ u_{i+1} < u_{i+2})
\end{multline*}
Analogously,
\begin{multline*}
\pi_t(d_{i-1}=1, u_{i+1} > u_{i+2})  
\\ = 
(1/2) \sum_{\sigma: \sigma_{i-1}=\sigma_{i}=\sigma_{i+1}=1} \pi_t(\sigma) P_{\sigma}(d_{i-1}=1~|~ u_{i+1} > u_{i+2}).
\end{multline*}
We claim that for any $s$ such that $s_{i-1}=s_{i}=s_{i+1}=1$, we have
\beql{eq-222}
P_{s}(d_{i-1}=1~|~ u_{i+1} < u_{i+2}) \ge P_{s}(d_{i-1}=1~|~ u_{i+1} > u_{i+2}).
\end{equation}
The proof of \eqn{eq-222} is by coupling the ranking and transmission realisations under the conditions 
$\{u_{i+1} < u_{i+2}\}$
and $\{u_{i+1} > u_{i+2}\}$, which we label as (a) and (b), respectively. The construction is as follows.
We use the process for producing a conditional ranking of this type, described in Section~\ref{sec-ranking-alg},
for conditions  $\{u_{i+1} < u_{i+2}\}$ and $\{u_{i+1} > u_{i+2}\}$, but we couple these processes.
Specifically, as long as neither $i+1$ nor $i+2$ is chosen yet, when the next node is being chosen
these two nodes are considered as a pair, which is picked with the probability twice greater than that a single node.
If/when the pair $(i+1,i+2)$ is chosen, the node $i+1$ or $i+2$ is picked {\em deterministically}, depending on whether the realisation is under condition (a) or (b), respectively. The process runs like this, and is stopped when (c1) node $i-1$ transmits or is deactivated, or (c2) the pair $(i+1,i+2)$ is chosen. In the former case, obviously, $u_{i-1}$ is same for conditions (a) and (b), so it remains to consider the case (c2) $\setminus$ (c1). Note that before the pair $(i+1,i+2)$ is chosen, nodes $i-1,i,i+1$ are active. If node $i+2$ was inactive, then under condition (a) node $i-1$ becomes an end node of an occupied segment (if we only consider occupied {\em active} nodes), and under condition (b) node $i-1$ is within the same occupied segment, but with nodes $i$ and $i+1$ appended ``on the right''; therefore, starting from this state of the access process, by Lemma~\ref{lem-friend}, the probability that node $i-1$ eventually transmits is higher under condition (a). Suppose node $i+2$ was active (and then necessarily occupied). Consider two subcases: (c2.1) not all nodes were occupied and active and (c2.2) all nodes were occupied and active. (For the line topology, we only need to consider the case (c2.1).) In case (c2.1), under condition (a) node $i-1$ becomes an end node of an occupied segment (if we only consider occupied {\em active} nodes), and under condition (b) node $i-1$ is within the same occupied segment, but with node $i$ appended ``on the right''; therefore, starting from this state of the access process, by Lemma~\ref{lem-friend}, the probability that node $i-1$ eventually transmits is higher under condition (a). In case (c2.2), under condition (a) node $i-1$ becomes an end node of the occupied segment of length $N-3$ (if we only consider occupied {\em active} nodes), and under condition (b) node $i-1$ is within the occupied segment of same length $N-3$, but with node $i-1$ being second from the end; therefore, starting from this state of the access process, by Lemma~\ref{lem-edge}, the probability that node $i-1$ eventually transmits is higher under condition (a). This completes the proof of 
claim \eqn{eq-222}.

Using \eqn{eq-222} we obtain
$$
\pi_t(d_{i-1}=1, u_{i+1} < u_{i+2}) \ge \pi_t(d_{i-1}=1, u_{i+1} > u_{i+2}).
$$
Given that
$$
\pi_t(d_{i-1}=1, u_{i+1} < u_{i+2}) + \pi_t(d_{i-1}=1, u_{i+1} > u_{i+2}) = \pi_t(d_{i-1}=1),
$$
we obtain \eqn{eq-111}.
$\Box$

\subsection{Proof of Theorem~\ref{thm:main}}

We are going to show that there exists $\epsilon>0$ such that for any fluid limit, at any regular point $t$
such that $\sum_{i=1}^N q_i(t) > 0$, 
\begin{equation} \label{eq:fluid_drift}
\sum_{i=1}^N q_i'(t) \le -\epsilon.
\end{equation}
This will prove stability, because, clearly, \eqn{eq:fluid_drift} implies \eqn{eq-fluid-key2}.

Consider first the case when $q_i(0) > 0$ for all $i$. In this case $\pi_t(s_i=1)=1$ for all $i$ and then, by symmetry,
$\pi_t(d_i=1) = C_N/N$ for all $i$. Then, 
 $q_i'(t) = \lambda - C_N/N < 0$ for all $i$ by Lemma \ref{lemma:25}, and \eqref{eq:fluid_drift} follows 
 (for any $\epsilon < C_N/N - \lambda$).


We now turn to the case when there is at least one $i$ with $q_i(t)=0$. 
A group of nodes $(k+1,k+2,\ldots,k+l)$ such that $q_k(t) = q_{k+l+1}(t) = 0$ and $q_{k+i}(t) >0$ for all $i=1,\ldots,l$, we will call a {\em positive group} of size $l$. We will show that 
\begin{equation} \label{eq:fluid_drift_group}
\sum_{i=1}^{l} q_{k+i}'(t) < -\epsilon_1 < 0,
\end{equation}
for any positive group $(k+1,k+2,\ldots,k+l)$ and, moreover, the constant $\epsilon_1$ in our estimates will depend only on the size $l$ of the group. Since there is only a finite number of possible values of $l$, \eqref{eq:fluid_drift} will follow.

We are going to present our proof of \eqn{eq:fluid_drift_group} for different values of $l$.

\begin{enumerate}

\item First consider the case $l=1$. Recall $\pi_t(s_{k+1}=1)=1$. Let 
$$
a = \pi_t(s_{k}=1, s_{k+2}=1)
$$ 
and let 
$$
b = \pi_t(s_{k-1}=1, s_{k+3}=1~|~s_{k}=1, s_{k+2}=1)
$$


From Lemma~\ref{lem-edge-triv} we immediately obtain
$$
\pi_t(d_{k+1} = 1~|~s_k=0 \bigcup s_{k+2}=0) \ge 1/2.
$$
And from Lemma \ref{lem-friend} we get
$$
\pi_t(d_{k+1} = 1~|~s_k=1, s_{k+2}=1, s_{k-1} = 0 \bigcup s_{k+3}=0) \ge 1/3
$$
and
$$
\pi_t(d_{k+1} = 1~|~s_k=1, s_{k+2}=1, s_{k-1} =1, s_{k+3}=1) \ge 179/420.
$$
Indeed, due to Lemma \ref{lem-friend}, in the first case the worst possible $s$ (appearing in the condition)
in terms of the smallest $P_s(d_{k+1}=1)$ is such that $k+1$ is the middle node in an occupied segment of length $3$, and in the second case - such that it is the middle node in an occupied segment of length $7$. The corresponding probabilities may either be calculated directly, or using \cite[Theorem 1]{Shneer2017}.

We obtain:
\begin{multline*}
q_{k+1}'(t) \le \lambda - 1/2(1-a) - a((1-b)(1/3)+b(179/420))  \\ = \lambda - 1/2 + a/2 - a(1/3 + b(13/140)) = \lambda - 1/2 + (1/2)a(1/3-b(13/70)).
\end{multline*}
Therefore, if 
\begin{equation} \label{eq:suff1}
a(1/3-b(13/70)) < \lambda/2, 
\end{equation}
then $q_{k+1}'(t) < \lambda - 1/2 + \lambda/4 < 0$, provided $\lambda < 2/5$, and \eqref{eq:fluid_drift_group} follows.

We can also write a different estimate of $q_{k+1}'(t)$:
\begin{align*}
q_{k+1}'(t) & = \lambda - \pi_t(d_k=0, d_{k+2} = 0) = \lambda - 1 + \pi_t(d_k=1 \bigcup d_{k+2}=1) 
\\ & =  \lambda - 1 + \pi_t(d_{k} =1) + \pi_t(d_{k+2}=1) - \pi_t(d_{k}=1, d_{k+2}=1) 
\\ & = \lambda -1+2\lambda - \pi_t(d_{k}=1, d_{k+2}=1) 
\\ & = 3\lambda - 1 - a \pi_t(d_{k}=1, d_{k+2}=1|s_{k} =1, s_{k+2} =1),
\end{align*}
and to prove \eqref{eq:fluid_drift_group} it is sufficient to show that
\begin{equation} \label{eq:suff2}
a \pi_t(d_{k}=1, d_{k+2}=1|s_{k} =1, s_{k+2} =1) > \lambda/2.
\end{equation}

To summarise, in order to prove \eqref{eq:fluid_drift_group} in this case, it is sufficient to prove either \eqref{eq:suff1} or \eqref{eq:suff2}.

Note that, thanks to Lemma \ref{lem-friend},
$$
\pi_t(d_{k}=1, d_{k+2}=1|s_{k} =1, s_{k+2} =1, s_{k-1} =1, s_{k+3} =1) \ge 1/5
$$
and
$$
\pi_t(d_{k}=1, d_{k+2}=1|s_{k} =1, s_{k+2} =1, s_{k-1} =0 \bigcup s_{k+3} = 0) \ge 3/8.
$$

Indeed, due to Lemma \ref{lem-friend}, in the first case the worst possible $s$ (appearing in the condition)
in terms of the smallest $P_s(d_k=1, d_{k+2}=1)$ is such that $k$ and $k+2$ are nodes $2$ and $4$ in an occupied segment of length $5$, and in the second case - such that they are nodes $2$ and $4$ in an occupied segment of length $4$. The corresponding probabilities are easy to verify directly.

We then have
\begin{align*}
& \pi_t(d_{k}=1, d_{k+2}=1|s_{k} =1, s_{k+2} =1) \\ & \ge (1/5)b + (3/8)(1-b) = 3/8 - b(7/40),
\end{align*}
and, in order to show \eqref{eq:suff2}, it is sufficient to show that
$$
a(3/8 - b(7/40)) > \lambda/2.
$$
Simple algebra shows that either the above or \eqref{eq:suff1} holds for any values of $a > 0$ and $b>0$.

\item Assume now that $l=2$. Note that $\pi_t(d_{k+1}+d_{k+2} =1~|~d_k=0) = 1$.
(When node $k$ is not transmitting, exactly one of nodes $k+1$ and $k+2$ will transmit.) Therefore,
\begin{align*}
q_{k+1}'(t) + q_{k+2}'(t) & = 2\lambda - \pi_t(d_k=1) \pi_t(d_{k+2}=1|d_k=1) - \pi_t(d_k=0) 
\\ & = 3\lambda -1 - \lambda \pi_t(d_{k+2}=1|d_k=1) \le (5/2)\lambda - 1 <0,
\end{align*}
because (from Lemma~\ref{lem-two-at-the-edge})
$$
\pi_t(d_{k+2}=1|d_k=1) \ge 1/2.
$$

\item Assume now that $l=3$ and note that in this case
\begin{align*}
& q_{k+1}'(t) + q_{k+2}'(t) + q_{k+3}'(t) 
\\ & = 3\lambda - (\pi_t(d_{k+1}=1) + \pi_t(d_{k+2}=1) + \pi_t(d_{k+3}=1)) \\
& = 3\lambda - (\pi_t(d_{k+1}=1) + \pi_t(d_{k+3}=1)) - (1 - \pi_t(d_{k+1}=1 \bigcup d_{k+3} = 1)) \\
& = 3\lambda -  1 - \pi_t(d_{k+1}=1, d_{k+3} = 1) \le 3\lambda-1-1/5 < 0
\end{align*}
if $\lambda < 2/5$, where the second-last inequality above follows from Lemma \ref{lem-friend}, again by
considering $s$ for which $\pi_t(d_{k+1}=1, d_{k+3} = 1)$ is the smallest, equal $1/5$.

\item Finally, consider the case $l \ge 4$. Let us first focus on nodes $k+1$ and $k+2$. Note that in our proof for the case $l=2$ we only used the fact that for the node to the left, node $k$, $\pi_t(d_k=1) = \lambda$. Therefore the proof may be repeated to show that
$$
q_{k+1}'(t) + q_{k+2}'(t) < 0.
$$
The same argument of course implies that
$$
q_{k+l-1}'(t) + q_{k+l}'(t) < 0.
$$
Note also that, due to Lemma \ref{lem-friend}, for any $3 \le i \le l-2$, $\pi_t(d_{k+i}=1) \ge 179/420$, as the ``worst '' occupancy state $s$ (among those with $s_j=1$ for $i-2\le j \le i+2$), in the sense of the smallest
$P_s(d_i=1)$, is the one where $i$ is the middle node within an occupied segment of length $7$ (and the corresponding probability may be calculated directly of from \cite[Theorem 1]{Shneer2017}).
As $179/420 > 2/5$, our proof is complete.

\end{enumerate}

\qed

\subsection{Proof of Theorem~\ref{thm:main_line}}

The proof of Theorem~\ref{thm:main_line} follows the same lines as that of Theorem~\ref{thm:main}.

Note that for all $N \ge 4$, $C_N = 1 + L_{N-3} \le L_N$ and therefore, due to Lemma~\ref{lemma:25}, $L_N/N \ge 2/5$ and
$$
\sum_{i=1}^N q_i'(t) = N \lambda - L_N < -\varepsilon
$$
with a positive $\varepsilon$, if $q_i(t) > 0$ for all $i$.

We can therefore consider positive groups of nodes bordering at least on one side an existing node $i$ with $q_i(t)=0$. Note that having no node on one side is beneficial for a group of $l \ge 2$ compared to having an existing node with $q_i(t)=0$. Therefore the proof of the negativity of the total drift of a group of $l \ge 2$ nodes carries over from Theorem~\ref{thm:main} without changes.

The proof in the case $l=1$ is unchanged again if the node $k+1$ under consideration is $3,\ldots,N-2$ as it relies only on considering nodes at a distance at most $2$ from the target one. If node $1$ (or node $N$) is the node under consideration, then, due to Lemma \ref{lem-edge-triv}, its drift is at most $\lambda - 1/2 < 0$. Therefore it remains to look at the case when the node under consideration is $2$ (the case of node $N-1$ is exactly the same by symmetry). We are going to show that the drift of node $2$ is negative in this case. The proof follows the same lines as the one given for the case $l=1$ in Theorem~\ref{thm:main}, with small changes in the worst cases for transmissions of certain (groups of) nodes. We provide the argument here for completeness.

Recall $\pi_t(s_{2}=1)=1$. Let 
$$
a = \pi_t(s_{1}=1, s_{3}=1)
$$ 
and let 
$$
b = \pi_t(s_{4}=1~|~s_{1}=1, s_{3}=1)
$$


From Lemma~\ref{lem-edge-triv} we immediately obtain
$$
\pi_t(d_{2} = 1~|~s_1=0 \bigcup s_{3}=0) \ge 1/2.
$$
Trivially,
$$
\pi_t(d_{2} = 1~|~s_1=1, s_{3}=1, s_{4}=0) = 1/3
$$
and from Lemma \ref{lem-friend} we get
$$
\pi_t(d_{2} = 1~|~s_1=1, s_{3}=1, s_{4}=1) \ge 11/30.
$$
Indeed, due to Lemma \ref{lem-friend}, in the last case the worst possible $s$ (appearing in the condition)
in terms of the smallest $P_s(d_{2}=1)$ is such that it is node $2$ in a segment of $5$ non-empty nodes. The corresponding probability may either be calculated directly, or using formulas from \cite{Shneer2017}.

We obtain:
\begin{multline*}
q_{2}'(t) \le \lambda - 1/2(1-a) - a((1-b)(1/3)+b(11/30))  \\ = \lambda - 1/2 + a/2 - a(1/3 + b(1/30)) = \lambda - 1/2 + (1/2)a(1/3-b(1/15)).
\end{multline*}
Therefore, if 
\begin{equation} \label{eq:suff1_1}
a(1/3-b(1/15)) < \lambda/2, 
\end{equation}
then $q_{2}'(t) < \lambda - 1/2 + \lambda/4 < 0$, provided $\lambda < 2/5$, and $q_2'(t) < 0$.

We can also write a different estimate of $q_{2}'(t)$:
\begin{align*}
q_{2}'(t) & = \lambda - \pi_t(d_1=0, d_{3} = 0) = \lambda - 1 + \pi_t(d_1=1 \bigcup d_{3}=1) 
\\ & =  \lambda - 1 + \pi_t(d_{1} =1) + \pi_t(d_{3}=1) - \pi_t(d_{1}=1, d_{3}=1) 
\\ & = \lambda -1+2\lambda - \pi_t(d_{1}=1, d_{3}=1) 
\\ & = 3\lambda - 1 - a \pi_t(d_{1}=1, d_{3}=1|s_{1} =1, s_{3} =1),
\end{align*}
and to prove that $q_2'(t) < 0$ it is sufficient to show that
\begin{equation} \label{eq:suff2_1}
a \pi_t(d_{1}=1, d_{3}=1|s_{1} =1, s_{3} =1) > \lambda/2.
\end{equation}

To summarise, in order to prove that $q_2'(t) < 0$ in this case, it is sufficient to prove either \eqref{eq:suff1_1} or \eqref{eq:suff2_1}.

Note that, thanks to Lemma \ref{lem-friend},
$$
\pi_t(d_{1}=1, d_{3}=1|s_{1} =1, s_{3} =1, s_{4} =1) \ge 3/8
$$
and
$$
\pi_t(d_{1}=1, d_{3}=1|s_{1} =1, s_{3} =1,  s_4=0) = 2/3.
$$

Indeed, due to Lemma \ref{lem-friend}, in the first case the worst possible $s$ (appearing in the condition)
in terms of the smallest $P_s(d_1=1, d_{3}=1)$ is such that $1$ and $3$ are nodes $1$ and $3$ in an occupied segment of length $4$.

We then have
$$
\pi_t(d_{1}=1, d_{3}=1|s_{1} =1, s_{3} =1) \ge (3/8)b + (2/3)(1-b) = 2/3 - b(7/24),
$$
and, in order to show \eqref{eq:suff2}, it is sufficient to show that
$$
a(2/3 - b(7/24)) > \lambda/2.
$$
Simple algebra shows that either the above or \eqref{eq:suff1_1} holds for any values of $a > 0$ and $0<b<1$.
\qed

\section{Conclusions, conjectures and further work} \label{sec:conclusion}

We considered a network of $N$ nodes arranged in a circle or a line where neighbouring nodes cannot transmit at the same time and each node's medium access is governed by a discrete-time CSMA protocol. Our main result shows that the system in either topology is stable for any $N\ge 4$ if the per-node arrival rate $\lambda < 2/5$. This provides an upper bound on the loss in throughput the standard (decentralised) CSMA algorithm suffers compared to the maximum possible throughput of $\lfloor N/2\rfloor /N$ that is achieved by an ideal centralised protocol.

The result is intuitive for the circle topology. Indeed, instability for a system of $N$ nodes on a circle may be easily proved if $\lambda > C_N/N$, where $C_N/N$ is the parking constant which is higher than $2/5$ and converges, as $N$ goes to infinity, to $1/2(1-e^{-2}) \approx 0.4323$. Instability holds because, if all nodes are active, then each of them receives a throughput of $C_N/N$. We conjecture that stability in fact holds if $\lambda < C_N/N$. This is an open question.

We obtain a bound of $2/5$ due to the fact that in the most technically difficult parts of the paper (see, e.g. cases $l=1$ and $l=3$ in the proof of the main result) we focus on a segment containing $5$ nodes and study the probabilities of various activation states and covariances between occupancies of different nodes. The value of $C_N/N$ is equal to $2/5$ in the case $N=5$, so our bound is tight in this case. It may be possible to obtain tight bounds for larger networks if one can find an efficient way of studying the correlations between occupancies and activations of a larger number of nodes. We believe however that this is cumbersome for networks with $N$ even slightly larger than $5$ and infeasible for very large networks. Different techniques may therefore be needed to prove our conjecture for an arbitrary $N$. We believe however that the techniques developed in this paper are nevertheless interesting in their own right. In particular, they may be applied to obtain sufficient conditions for stability in networks with more general topologies. They might also become a part of different and/or more general techniques for the analysis of non-monotone processes.

Our result is {\em not} intuitive in the case of the line topology because, if all nodes are active, some receive a throughput of less than $2/5$. This illustrates how intricate the system behaviour is and how the analysis of unsaturated systems is difficult even in such simple topologies. The exact stability condition, even for a homogeneous system, is an open problem for the line topology; this condition is even not easy to conjecture.

We are currently also working on more difficult multi-hop networks where each message has a source and a destination and may therefore need to be transmitted by several nodes in the network. For a line we consider traffic arriving at node $1$ and requiring to reach node $N$. Each message thus needs to be transmitted by every node in turn. We are interested in both stability and end-to-end throughput of the system.

For a circle one can define a multi-hop network in, for instance, the following way. Fix a constant $m \ge 1$ and assume that each node gets on average $\lambda/m$ new packets per time slot. Assume that medium-access competition is done in the same way as for the single-hop system but, upon a successful transmission from node $i$, a message leaves the system with probability $1/m$ and goes to the queue of node $i+1$ with probability $1-1/m$. It is easy to show that the average nominal traffic load for any node is $\lambda$ and we conjecture that $\lambda < C_N/N$ is sufficient for stability for any $m \ge 1$. As $m \to \infty$ and $N \to \infty$ the model on a circle may serve as an approximation for the behaviour of a large line segment far away from source.

We can in fact prove the above multi-hop stability conjecture for $N=4$ using a simplified version of the argument of the proof of Theorem \ref{thm:main}. The proof however heavily relies on the specific structure of the system of $4$ nodes in that at every time slot there are, essentially, at most two possible outcomes of the medium-access competition. This proof does not generalise to larger values of $N$. Stability for all $N$ and all $m \ge 1$ is a challenging and exciting question for further study.

\section*{Acknowledgements}
We are grateful to the anonymous referee and associate editor for their careful reading of our paper and their comments and suggestions.

\section*{Appendix}

{\bf Proof of Lemma \ref{lemma:25}.} We are going to show that the sequence $C_n/n$ is non-increasing for even values of $n$ and non-decreasing for odd values of $n$. Therefore, for all even values of $n \ge 4$,
$$
\frac{C_n}{n} \ge \lim_{n \to \infty} \frac{C_n}{n} = \frac{1}{2}(1-e^{-2}) > 2/5,
$$
and for all odd values of $n \ge 5$
$$
\frac{C_n}{n} \ge \frac{C_5}{5} = 2/5.
$$
Using the expression for $L_k$ from \cite[Corollary 2]{Shneer2017}, we can write
\begin{multline*}
\frac{C_n}{n} = \frac{1+L_{n-3}}{n} = \cfrac{1+\sum\limits_{k=1}^{n-3}(-1)^{k+1} \cfrac{2^{k-1}}{k!} (n-2-k)}{n} \\
= \sum\limits_{k=1}^{n-3}(-1)^{k+1} \frac{2^{k-1}}{k!} + \frac{1}{n}\left(1 - \sum\limits_{k=1}^{n-3}(-1)^{k+1} \frac{2^{k-1}}{k!} (k+2) \right)\\
= \sum\limits_{k=1}^{n-3}(-1)^{k+1} \frac{2^{k-1}}{k!} + \frac{1}{n}\left(1 - \sum\limits_{k=1}^{n-3}(-1)^{k+1} \frac{2^{k-1}}{(k-1)!} - \sum\limits_{k=1}^{n-3}(-1)^{k+1} \frac{2^k}{k!} \right)\\
= \sum\limits_{k=1}^{n-3}(-1)^{k+1} \frac{2^{k-1}}{k!} + \frac{1}{n}\left(1 - \sum\limits_{l=0}^{n-4}(-1)^{l+2} \frac{2^l}{l!} - \sum\limits_{k=1}^{n-3}(-1)^{k+1} \frac{2^k}{k!} \right) \\
= \sum\limits_{k=1}^{n-3}(-1)^{k+1} \frac{2^{k-1}}{k!} + \frac{1}{n}\left(1-1-(-1)^{n-2}\frac{2^{n-3}}{(n-3)!} \right) \\
= \sum\limits_{k=1}^{n-3}(-1)^{k+1} \frac{2^{k-1}}{k!} - \frac{1}{n}(-1)^{n}\frac{2^{n-3}}{(n-3)!}.
\end{multline*}

Consider now
\begin{multline*}
\frac{C_{n+2}}{n+2} - \frac{C_n}{n} 
\\ = (-1)^{n-1}\frac{2^{n-3}}{(n-2)!} + (-1)^{n}\frac{2^{n-2}}{(n-1)!} - \frac{1}{n+2}(-1)^{n}\frac{2^{n-1}}{(n-1)!} + \frac{1}{n}(-1)^{n}\frac{2^{n-3}}{(n-3)!} \\
= (-1)^{n}\frac{2^{n-3}}{(n-3)!}\left(-\frac{1}{n-2} + \frac{2}{(n-2)(n-1)} - \frac{4}{(n+2)(n-1)(n-2)} + \frac{1}{n} \right)
\\ = (-1)^{n}\frac{2^{n-3}}{(n-3)!}\left(-\frac{1}{n-2} + \frac{2n}{(n+2)(n-2)(n-1)} + \frac{1}{n} \right)
\\ = (-1)^{n}\frac{2^{n-3}}{(n-3)!}\left(- \frac{n+1}{(n+2)(n-1)} + \frac{1}{n} \right) 
= (-1)^{n}\frac{2^{n-3}}{(n-3)!}\left(- \frac{2}{n(n+2)(n-1)} \right),
\end{multline*}
and this completes the proof.
\qed

\end{document}